\documentclass{article}
\usepackage{spconf,amsmath,graphicx}

\usepackage{url}

\usepackage{array,multirow,graphicx}
\usepackage{float}
\usepackage{graphicx}
\usepackage{xparse,xfp}
\usepackage{amsmath, amssymb}
\usepackage{cite}
\usepackage{tikz}
\usepackage{float}
\usepackage[capitalize]{cleveref}
\usetikzlibrary{shapes,arrows, decorations}
\usepackage{pifont}
\usepackage{microtype}      
\usepackage{enumitem}
\usepackage{subcaption}

\tikzstyle{specialblock} = [draw, ultra thick, fill=blue!20, rectangle, 
    minimum height=3em, minimum width=4em]
\tikzstyle{block} = [draw, fill=lightgray, rectangle, 
    minimum height=3em, minimum width=4em]
\tikzstyle{sum} = [draw, fill=white, circle, node distance=1cm]
\tikzstyle{prod}   = [circle, minimum width=8pt, draw, inner sep=0pt, path picture={\draw (path picture bounding box.south east) -- (path picture bounding box.north west) (path picture bounding box.south west) -- (path picture bounding box.north east);}]
\tikzstyle{sumt}   = [circle, minimum width=8pt, draw, inner sep=0pt, path picture={\draw (path picture bounding box.east) -- (path picture bounding box.west) (path picture bounding box.south) -- (path picture bounding box.north);}]
\tikzstyle{input} = [coordinate]
\tikzstyle{output} = [coordinate]
\tikzstyle{pinstyle} = [pin edge={to-,thin,black}]
\tikzset{
tmp/.style  = {coordinate}, 
dot/.style = {circle, minimum size=#1,
              inner sep=0pt, outer sep=0pt},
dot/.default = 6pt 
}

\title{Resource-Efficient Separation Transformer}

\name{
\parbox{\linewidth}{\centering
{Luca Della Libera\thanks{$^{*}$Equal contribution.}$^{*1}$}, Cem Subakan$^{*2,1,3}$, Mirco Ravanelli$^{1,3}$,\\Samuele Cornell$^{4}$, Frédéric Lepoutre$^5$, François Grondin$^6$}}

\address{
\parbox{\linewidth}{\centering
$^1$Concordia University, $^2$Université Laval, $^3$Mila-Quebec AI Institute,\\$^4$Università Politecnica delle Marche, $^5$Soundskrit Inc., $^6$Université de Sherbrooke}}

\begin{document}
\ninept

\maketitle

\begin{abstract}
Transformers have recently achieved state-of-the-art performance in speech separation. These models, however, are computationally demanding and require a lot of learnable parameters.
This paper explores Transformer-based speech separation with a reduced computational cost. Our main contribution is the development of the Resource-Efficient Separation Transformer (RE-SepFormer), a self-attention-based architecture that reduces the computational burden in two ways. First, it uses non-overlapping blocks in the latent space. Second, it operates on compact latent summaries calculated from each chunk.
The RE-SepFormer reaches a competitive performance on the popular WSJ0-2Mix  and WHAM! datasets in both causal and non-causal settings. Remarkably, it scales significantly better than the previous Transformer-based architectures in terms of memory and inference time, making it more suitable for processing long mixtures.
\end{abstract}


\newcommand{\mixture}{x}
\newcommand{\ldim}{F}
\newcommand{\len}{T}
\newcommand{\llen}{T'}
\newcommand{\nspk}{N_s}
\newcommand{\nsepf}{M}
\newcommand{\chnksize}{C}
\newcommand{\hopsize}{H}
\newcommand{\numchnks}{N_c}
\newcommand{\numintra}{Nintra}
\newcommand{\numinter}{Ninter}
\newcommand{\trlat}{e}

\section{Introduction}
In recent years, deep learning has become more and more computationally demanding. The current trend consists of improving the performance of Deep Neural Networks (DNNs) through ever-larger models crunching ever-larger amounts of data. In natural language processing (NLP), this tendency led to large language models such as GPT3 \cite{gpt3}, PaLM \cite{palm}, Megatron \cite{shoeybi2020megatronlm} and many others. Similarly, large neural networks like wav2vec2.0 \cite{wav2vec2.0} and HuBERT \cite{hubert} have gained popularity for speech processing. 
Large neural models, however, are energy-demanding, causing high inference costs (with considerable CO2 emissions \cite{Parcollet2021TheEA}) that limit their widespread adoption in production systems. Moreover, such models cannot process users' data on the device, thus raising privacy concerns \cite{privacy}.
Efficient deep learning has been the object of recent research efforts. Approaches such as distillation \cite{hinton-distillation}, neural network pruning \cite{reed1993, molchanov2016}, and binarization/quantization \cite{kim2016, hubara2016} have been explored.
In the context of speech processing, various efficient methods for speech enhancement/separation \cite{phasen, dcrnn, luo2018convtasnet, luo2020dualpath, tzinis2020sudo, subakan2022onusing, skim}, keyword spotting, language identification \cite{mazzawi19_interspeech}, emotion recognition \cite{wu2022}, and automatic speech recognition \cite{hu2020} have been proposed.\looseness=-1

Among all the popular neural models, Transformers \cite{vaswani2017} are particularly difficult to make computationally efficient due to their quadratic memory bottleneck and the high number of parameters that they typically require. Revised Transformer-based architectures, which relax the quadratic memory requirement, \cite{tay2020efficient} such as the Linformer \cite{wang2020linformer}, Longformer \cite{beltagy2020longformer}, and Reformer \cite{kitaev2020reformer} have been proposed recently. Efficient Transformers have been studied for speech processing tasks as well, such as speech recognition \cite{kim2022squeezeformer}, enhancement, and separation \cite{subakan2022onusing}.
These previous works, however, do not consider causal models suitable for on-device real-time applications. 

In this paper, we propose a novel small-footprint speech separation model built upon the SkiM framework \cite{skim}, called Resource-Efficient Separation Transformer (\emph{RE-SepFormer}), that represents a lightweight alternative to the recently-proposed SepFormer. Unlike the SepFormer, the RE-SepFormer uses non-overlapping chunks in the latent space and, therefore, it reduces by half the number of chunks to process (if we consider the default overlap rate of 50\%).
Moreover, we further reduce the computations by using a special mechanism called \emph{Memory Transformer}. The Memory Transformer operates over a summary representation calculated from whole chunks rather than attending every single element of the chunk independently. 

We conducted the experimental validation on the popular WSJ0-2Mix dataset. To assess our model in more realistic conditions, we also considered the WHAM! \cite{wichern2019wham} dataset, which contains mixtures corrupted by non-stationary environmental noise. In addition to the non-causal offline scenario we provide experimental evidence in a real-time low-latency scenario by considering causal versions of the RE-SepFormer. 
In detail, our contributions are the following:
\begin{itemize}[leftmargin=0.40cm]
 \item We show that the RE-SepFormer reaches a competitive performance on the popular WSJ0-2Mix and  WHAM! datasets in both causal and non-causal settings. 

\item With the RE-SepFormer, we achieve a 3x reduction of the parameters and 11x reduction of the multiply-accumulate operations (MACs) per second over the standard SepFormer.

\item The RE-SepFormer is extremely parallelizable, and therefore highly suitable for GPU inference. It scales significantly better than the previous Transformer-based architectures in terms of memory and inference time, making it more suitable for processing long mixtures.\looseness=-1
\end{itemize}

\section{RE-SepFormer}
\subsection{Time-Domain Masking Architecture}
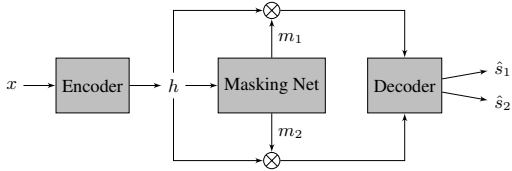
\begin{figure}[t]
\centering
    \resizebox{7.0cm}{!}{
    \begin{tikzpicture}[auto, node distance=2.2cm,>=latex']
        \hspace*{-0.02\linewidth}
        \node [draw=none, fill=none] (input) {$\mixture$ };
        \node [block, right of=input, fill, xshift=-0.8cm ] (encoder) {Encoder};  
        \node [draw=none, fill=none, right of=encoder, xshift=-0.8cm] (latent) {$h$};
        \node [block, right of=latent, xshift=-0.5cm] (masknet) {Masking Net};
        \node [prod, above of=masknet,yshift=-0.9cm] (prod1) {}; 
        \node [prod, below of=masknet,yshift=0.9cm] (prod2) {}; 
        \node [block, right of=masknet, xshift=0.1cm] (decoder) {Decoder};
        \node [draw=none, fill=none, right of=decoder, yshift=0.3cm, xshift=-0.5cm] (out1) {$\hat{s}_1$};
        \node [draw=none, fill=none, right of=decoder, yshift=-0.3cm, xshift=-0.5cm] (out2) {$\hat{s}_2$};
        
        \draw [->] (input) -- (encoder);
        \draw [->] (encoder) -- (latent);
        \draw [->] (latent) -- (masknet);
        \draw [->] (latent) |-  (prod1);
        \draw [->] (latent) |-  (prod2);
        \draw [->,swap] (masknet) -- node {$m_1$} (prod1);
        \draw [->] (masknet) -- node {$m_2$} (prod2);
        \draw [->] (prod1) -| (decoder);
        \draw [->] (prod2) -| (decoder);
        \draw [->] (decoder) -- (out1);
        \draw [->] (decoder) -- (out2);
    \end{tikzpicture}
    }
    \vspace{-.1cm}
\caption{An high-level description of the masking-based source separation pipeline: the encoder learns a latent representation $h$ from the input mixture $x$. The masking network then estimates the optimal masks $m_1$ and $m_2$ to separate the sources in the mixture. Finally, the decoder reconstructs the sources from the masked representations.\vspace{-0.3cm}}
\label{fig:maskingpipeline}
\end{figure}

The overall architecture of the RE-SepFormer is shown in \cref{fig:maskingpipeline}.
Similar to other popular architectures such as Conv-TasNet \cite{luo2018convtasnet}, Dual-Path RNN \cite{luo2020dualpath}, and SepFormer \cite{subakan2020attention}, our model is based on learned-domain masking. The input mixture $x \in \mathbb R^\len$ is first passed through an encoder, which
outputs a latent representation $h \in \mathbb R^{\llen \times \ldim}$ using a strided convolutional layer:
\begin{align}
    h = \text{ReLU}(\text{conv1d}(x)).
\end{align}
The masking network takes in the representation $h$, and produces the masks $m_1, m_2 \in \mathbb R^{\llen \times \ldim}$. We consider here a model with two sources without loss of generality. 
The decoder is a transposed convolutional layer that reconstructs the time-domain signals $\hat{s}_1$ and $\hat{s}_2$ from the masked latent representations:
\begin{align}
    \widehat {s}_k = \text{conv1d-transpose}(m_k * h). 
\end{align}

\subsection{The Masking Network}

\tikzstyle{chunks} = [draw, thick, fill=blue!10, rectangle, 
    minimum height=4.2cm, minimum width=2cm] 
    
    \tikzstyle{timeblock} = [draw, thick, fill=blue!10, rectangle, 
    minimum height=1.9em, minimum width=17cm]
    \newcommand{\septime}{1}
    \newcommand{\yshift}{7.3}

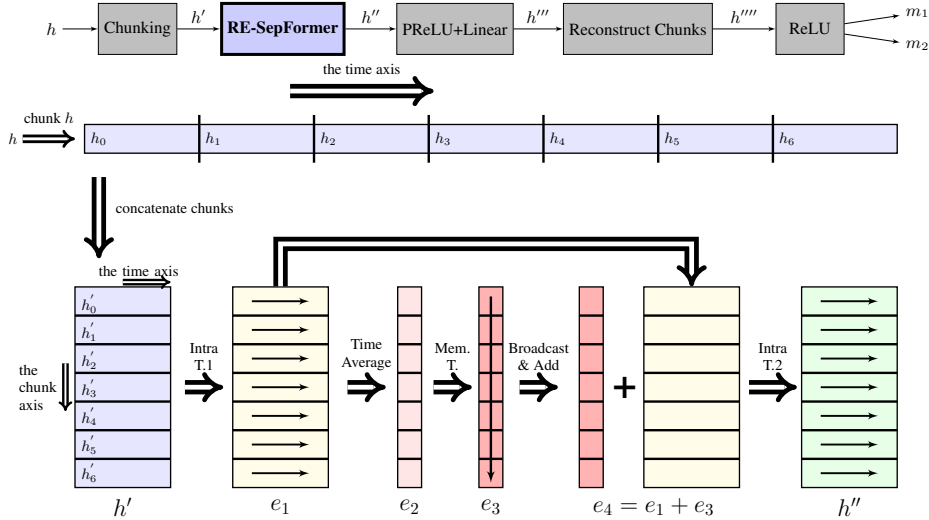
\begin{figure*}[t!]
\centering
    \newcommand{\sepfig}{0.8}
    \resizebox{12.0cm}{!}{
    \begin{tikzpicture}[auto, node distance=2.0cm,>=latex']
        \node [draw=none, fill=none] (h) {$h$};
        \node [block, right of=h, fill, xshift=-.4 cm] (2) {Chunking};  
        \node [specialblock, right of=2,  xshift=0.7 cm] (3) {\textbf{RE-SepFormer}};  
        \node [block, right of=3, xshift=1.2 cm] (4) {PReLU+Linear};
        \node [block, right of=4, xshift=1.4 cm] (5) {Reconstruct Chunks};
        \node [block, right of=5, xshift=1.2 cm] (6) {ReLU};
        \node [draw=none, fill=none, right of=6, yshift=0.3cm] (out1) {$m_1$};
        \node [draw=none, fill=none, right of=6, yshift=-0.3cm] (out2) {$m_2$};
        
        \draw [->] (h) -- (2); 
        \draw [->] (2) -- node {$h'$} (3); 
        \draw [->] (3) -- node {$h''$} (4); 
        \draw [->] (4) -- node {$h'''$} (5); 
        \draw [->] (5) -- node {$h''''$} (6); 
        \draw [->] (6) -- (out1); 
        \draw [->] (6) -- (out2); 
    \end{tikzpicture}
    }
    
    \resizebox{12.0cm}{!}{
    \begin{tikzpicture}[auto, node distance=1.0cm,>=latex']
    \hspace*{-0.04\linewidth}
        \node [xshift=-7.8cm, yshift=0cm] (hshow) {$h$};
        \draw  [line width=0.6mm, -implies,double, double distance=.6mm] (hshow)-- node (taxis) [yshift=0.2cm] {chunk $h$} (-6.5, 0);  
        
        \node [draw=none, fill=none] (h) {$h$};
        \node [timeblock, right of=h, fill, xshift=1.2cm] (asd) {};  
        
        \foreach \x in {0,...,5}{
            \draw [line width=0.5mm](-\x*2.4 + 8.1, -0.5) -- (-\x*2.4 + 8.1, 0.5);
        }
       
        \foreach \x in {0,...,6}{
            \node [xshift=\fpeval{-6.0 + \x*2.4 } cm](h\x) {$h_{\fpeval{\x}}$};
        }
        
        \newcommand{\tsh}{-0.5}
        \newcommand{\tshs}{-1.20}
        \node [chunks, below of=h, yshift=-4.2cm, xshift=-5.5cm, minimum width=2.0cm] (ch1) {};
        \foreach \y in {1,...,6}{
            \draw [line width=0.5mm](-4 + \tsh, 0.6*\y - \yshift) -- (-6 + \tsh, 0.6*\y-\yshift);
        }
        \foreach \y in {0,...,6}{
            \node [yshift=\fpeval{-0.59*\y-3.45} cm, xshift=\fpeval{\tsh + -5.7}cm](h\y) {$h^{'}_{\y}$};
        }
        
        \node [chunks, right of=ch1, xshift=2.3cm, fill=yellow!10] (ch2) {};
        \foreach \y in {1,...,6}{
           \draw [line width=0.5mm](-2 + \tshs, 0.6*\y - \yshift) -- (-0+\tshs, 0.6*\y-\yshift);
            
        }
        \foreach \y in {1,...,7}{
            \draw [->, line width=0.4mm] (-1.6 + \tshs, 0.6*\y - \yshift -0.3) -- (-0.4 + \tshs, 0.6*\y-\yshift -0.3);
        }
        
        \node [chunks, right of=ch2, xshift=1.7cm, fill=red!10, minimum width=.5cm] (ch3) {};
        \foreach \y in {1,...,6}{
            \draw [line width=0.5mm](1.450 + \tshs, 0.6*\y - \yshift) -- (1.950 + \tshs, 0.6*\y-\yshift);
        }
        
        \node [chunks, right of=ch3, xshift=0.7cm, fill=red!30, minimum width=.5cm] (ch4) {};
        \foreach \y in {1,...,6}{
            \draw [line width=0.5mm](3.15 + \tshs, 0.6*\y - \yshift) -- (3.65 + \tshs, 0.6*\y-\yshift);
        }
        \foreach \x in {0}{
            \draw [->, line width=0.5mm, color=black](\x*0.4 + \tshs + 3.4, -3.3) -- (\x*0.4 + 3.4 + \tshs, -7.2);
        }
        
         \node [chunks, right of=ch4, xshift=1.1cm, fill=red!30, minimum width=.5cm] (ch5) {};
         \foreach \y in {1,...,6}{
            \draw [line width=0.5mm](5.25 + \tshs, 0.6*\y - \yshift) -- (5.75 + \tshs, 0.6*\y-\yshift);
        }
        
         \node [chunks, right of=ch5, xshift=1.1cm, fill=yellow!10] (ch6) {};
         \foreach \y in {1,...,6}{
           \draw [line width=0.5mm](6.6 + \tshs, 0.6*\y - \yshift) -- (8.6+\tshs, 0.6*\y-\yshift);
        }
         
         \node [chunks, right of=ch6, xshift=2.3cm, fill=green!10] (ch7) {};
        \foreach \y in {1,...,6}{
           \draw [line width=0.5mm](9.9 + \tshs, 0.6*\y - \yshift) -- (11.9+\tshs, 0.6*\y-\yshift);
        }
        \foreach \y in {1,...,7}{
            \draw [->, line width=0.4mm] (10.3 + \tshs, 0.6*\y - \yshift -0.3) -- (11.5 + \tshs, 0.6*\y-\yshift -0.3);
        }

        \draw [line width=1mm, -implies,double, double distance=1mm] (-2,1)-- node (taxis) [yshift=0.2cm] {the time axis} (1,1); 
        
        \draw [line width=1mm, -implies,double, double distance=1mm] (-6,-0.8)-- node (taxis) [yshift=0.2cm, xshift=0.2cm] {concatenate chunks} (-6, -2.5); 
        
        \draw [line width=.5mm, -implies,double, double distance=0.4mm, yshift=0.3cm, xshift=0.3cm] (-6.5+\tsh,-5)-- node (taxis) [xshift=-1.1cm, text width=1cm] {the chunk axis} (-6.5+\tsh, -6); 
        
        \draw [line width=.5mm, -implies,double, double distance=0.4mm] (-5+\tsh,-3)-- node (taxis) [xshift=-0.0cm, text width=2cm] {the time axis} (-4+\tsh, -3); 
        
        \draw [line width=1mm, -implies,double, double distance=1mm, yshift=0.3cm] (\fpeval{-3 + \tshs}, -5.5)-- node (taxis) [yshift=0.2cm, align=center] {Intra \\ T.1} (-2.2 + \tshs, -5.5);         
        
        \draw [line width=1mm, -implies,double, double distance=1mm, yshift=0.3cm] (0.4 + \tshs, -5.5)-- node (taxis) [align=center, yshift=0.2cm] {Time \\ Average} (1.2 + \tshs, -5.5); 
        
        \draw [line width=1mm, -implies,double, double distance=1mm, yshift=0.3cm] (2.2 + \tshs, -5.5)-- node (taxis) [yshift=0.2cm, align=center] {Mem. \\ T.} (3.0 + \tshs, -5.5); 
        
        \draw [line width=1mm, -implies,double, double distance=1mm, yshift=0.3cm, xshift=.2cm] (3.8 + \tshs, -5.5)-- node (taxis) [yshift=0.2cm, align=center] {Broadcast \\ \& Add} (4.6 + \tshs, -5.5); 
        
        \node [xshift=\fpeval{6.2+\tshs}cm, yshift=-5.2cm] (plus) {\Huge{\textbf{+}}};  
        
        \draw [line width=1mm, -implies,double, double distance=1mm, yshift=0.3cm] (8.8+\tshs, -5.5)-- node (taxis) [yshift=0.2cm, align=center] {Intra \\ T.2} (9.8 +\tshs, -5.5); 
        
        \node [draw=none, above of=ch4, yshift=2cm] (ch4a) {};
        \draw [->, line width=1mm, -implies,double, double distance=1mm] (ch2) |-(ch4a.center)-| (ch6);
        
        \node [below of=ch1, yshift=-1.5cm] () {\Large{$h'$}};
        \node [below of=ch2, yshift=-1.5cm] () {\Large{$\trlat_1$}};
        \node [below of=ch3, yshift=-1.5cm] () {\Large{$\trlat_2$}};
        \node [below of=ch4, yshift=-1.5cm] () {\Large{$\trlat_3$}};
        \node [below of=plus, yshift=-1.5cm] () {\Large{\; \; \; \; \;$\trlat_4 = \trlat_1 + \trlat_3$}};
        \node [below of=ch7, yshift=-1.5cm] () {\Large{$h''$}};
    \end{tikzpicture}
    }
    \vspace{-.25cm}
    
    \caption{
    (\textbf{Top}) The architecture of the masking network. (\textbf{Bottom}) The Resource-Efficient SepFormer module: (1) the latent representation $h$ is chunked to get $h_0'$, $h_1'$, $\dots$, $h_{\numchnks}'$ (2) the IntraTransformer is applied to all of the chunks independently (3) the output is averaged over the time dimension and passed through the memory Transformer (4) the resulting vector is added to the output of the IntraTransformer with broadcasting over the time axis (5) the resulting tensor is passed through another IntraTransformer to obtain the final output $h''$.} 
    
\label{figure:maskingblockdiagram}
\end{figure*}

The architecture of the masking network is depicted in \cref{figure:maskingblockdiagram} (top). 
Firstly, the input representation is split into temporal chunks. We denote this tensor with $h' \in \mathbb R^{\chnksize \times \numchnks \times \ldim}$, where $\chnksize$ is the size of each chunk, and $\numchnks$ is the resulting number of chunks. As opposed to the standard SepFormer, we use non-overlapping chunks to reduce the amount of computations.
Then, the RE-SepFormer processes the chunks $h'$ and provides a latent representation $h'' \in \mathbb R^{\chnksize \times \numchnks \times \nspk \times \ldim}$  (where $N_s$ denotes the number of sources). The tensor $h'' $ is further transformed by a PReLU activation function and a linear layer. 
Finally, the chunking is undone by simply concatenating the chunks on the original time axis. We denote the resulting tensor with $h'''' \in \mathbb R^{\llen \times \nspk \times \ldim}$. The final masks are estimated by passing this tensor through a ReLU non-linearity.

\subsection{Resource Efficient SepFormer}
The RE-SepFormer block is detailed in \cref{figure:maskingblockdiagram} (bottom). This module includes three main components: the \textit{IntraTransformer1}, the \textit{MemoryTransformer}, and the \textit{IntraTransformer2}. 
The IntraTransformer1 is applied to the time axis of all of the chunks and generates a tensor $\trlat_1 \in \mathbb R^{\chnksize \times \numchnks \times \ldim}$. The goal of this block is to process short-term temporal information.
We then compute a summary representation $\trlat_2 \in \mathbb R^{\numchnks \times \ldim}$ by averaging this tensor over the time axis. The intuition behind this operation is that the average over the time-axis of a latent representation can provide enough high-level contextual information to embed longer-term dependencies. Working with a summary vector is much more computationally convenient than operating on the full tensor $\trlat_1$ as done in the original SepFormer. This saves significant amounts of computations. 
 
This summary representation $\trlat_2$ is then fed into the {Memory Transformer}. The latter is applied to the chunk axis and produces a representation $\trlat_3 \in \mathbb R^{\numchnks \times \ldim}$ that models long-term dependencies across chunks. The $\trlat_3$ tensor is then added element-wise to $\trlat_1$ (with broadcasting over the time axis). The resulting $\trlat_4$ tensor is particularly rich as it incorporates both short and long-term dependencies. Note also that this operation implicitly adds a gradient shortcut in the architecture, contributing to making the architecture easier to train and more robust against vanishing gradient issues.

Finally, we provide more capacity to the model by feeding $\trlat_4$ into another IntraTransformer2 operating on the time axis. This generates the tensor $h'' \in \mathbb R^{\chnksize \times \numchnks \times \nspk \times \ldim}$, that is the output of the RE-SepFormer block.

\section{Experimental Setup}
\subsection{Datasets} 
We provide experimental evidence on the popular WSJ0-2Mix dataset \cite{hershey2015deep}, which is a standard benchmark largely adopted in speech separation. The training, validation, and test partitions consist of 30, 10, and 5 hours of speech. The mixtures are created by randomly mixing utterances using random relative gains between 0 dB and 5 dB. In the training and test sets, different speakers are used.
We also assess our models on the WHAM! corpus \cite{wichern2019wham}, which adds non-stationary noises such as background noise from restaurants, bars, and parks with different SNRs to the mixtures.

\subsection{Models}
We compare the RE-SepFormer against popular speech separation methods (e.g., TasNet, Conv-TasNet, Dual-Path RNN, etc.).
Moreover, we compare our model with SkiM \cite{skim}, which is an RNN-based model recently proposed to perform computationally-efficient speech separation. We obtained the implementation of SkiM \cite{skim} from the ESPnet toolkit \cite{espnet} and used the default parameter setting reported in the original paper. We also compare against SepFormer-Light that has 6.4M parameters down from 25.7M. For this reduction, we used 128 encoder outputs and 512 dimensions in the feed-forward layers of the Transformer. The implementation of RE-SepFormer is available in the SpeechBrain \cite{speechbrain} GitHub repository\footnote{\url{https://github.com/speechbrain/speechbrain/tree/develop/recipes/WSJ0Mix/separation}}.

We consider both the causal and non-causal versions of these models.
The causal version of the RE-SepFormer uses triangular matrices in the attention mechanisms that prevent attending to future time steps.
For the RE-SepFormer, we used 128 convolutional filters in the convolutional encoder. We used 8 Transformer layers for the Intra and Memory Transformers with 8 parallel attention heads and 1024 dimensional positional feed-forward layers. The chunk size is set to 150 samples.\looseness=-1

\subsection{Training Details} 
\label{subsec:arc-tr-details}
The model is trained with the permutation invariant SI-SNR loss \cite{luo2018convtasnet, kolbaek2017multitalker, le2019sdr} and the parameters are updated with the Adam optimizer \cite{kingma2017adam}.
We halve the learning rate after epoch 85 if the performance on the validation set does not improve for 3 consecutive epochs.
We train the model with dynamic mixing (DM) \cite{tzinis2019twostep, zeghidour2020wavesplit}, which generates new mixtures on-the-fly by randomly mixing clean utterances. 
For more details, please refer to the reference WSJ0-Mix recipe available on SpeechBrain \cite{speechbrain}.\looseness=-1

\section{Results}
\subsection{From SepFormer to RE-SepFormer}
\label{subsec:sep2resep}
In \cref{table:sep2resep} we show the impact of the architectural changes done to derive the RE-SepFormer from the original SepFormer in terms of performance, number of parameters, and MACs using the standard WSJ0-2Mix dataset (non-causal setting).

The first modification is the use of non-overlapped chunks.
From \cref{table:sep2resep}, it emerges that there is a significant performance drop when we adopt this change (see first two lines). The SDRi, for instance, drops from 22.4 dB (standard SepFormer) to 16.2 dB. On the other hand, we significantly reduce the MACs by 2.5x. 

The other proposed intervention is the summary representation (time average) processed by a Memory Transformer. This operation drastically reduces the number of parameters (3.2x reduction) and the MACs (11x reduction). Interestingly, this modification not only does not deteriorate the separation performance but even yields a slight improvement, possibly due to the fact that the averaging operation better promotes continuity between the latent chunks. 
We believe that the RE-SepFormer is particularly interesting for small-footprint devices. Even though the performance drop is not negligible compared to the standard SepFormer, the model still provides very high-quality speech separation (SDRi up to 18.9 dB) with a drastic reduction of computational resources.\looseness-1%

\begin{table}[t]
\vspace{-0.4cm}
\centering
\caption{Comparison of SepFormer and RE-SepFormer on WSJ0-2Mix, non-causal setting.} 
\label{table:sep2resep}
\resizebox{8.7cm}{!}{
\begin{tabular}{c|c|c|c|c|c|c}
\textbf{Model} & \textbf{Overlap}  & \textbf{Avg} & \textbf{SI-SNRi (dB)} & \textbf{SDRi (dB)} & \textbf{\#Params (M)} & \textbf{GMACs/s} \\
\hline  
SepFormer & 50\%  & $\times$  & 22.3  & 22.4 & 25.7 & 69.6   \\ \hline
SepFormer & 0\%  & $\times$ & 15.9  & 16.2 & 25.7 & 28.0   \\ \hline
RE-SepFormer & 0\%  & $\checkmark$ & 18.6  & 18.9  & 8.0 & 6.3   \\ \hline
\end{tabular}
}
\vspace{-0.4cm}
\end{table}

\subsection{Comparison with SkiM}
\label{subsec:comparisonskim}
\cref{table:resepvsskim} compares the performance of the RE-SepFormer with SkiM \cite{skim} on WSJ0-2Mix and WHAM! in causal and non-causal settings.
SkiM \cite{skim} is a recently proposed model for efficient speech separation. It uses RNNs with non-overlapping chunks, making it a natural benchmark.
For a fair comparison, we use dynamic mixing and the same kernel size in the convolutional layers ($kernelsize$ = 16) for both models.\looseness=-1

As shown in \cref{table:resepvsskim}, the RE-SepFormer outperforms SkiM in three of the four tested conditions: it provides better performance on the WSJ0-2Mix dataset (both causal and non-causal settings) and on the WHAM! corpus (causal setting). SkiM, on the other hand, slightly outperforms the RE-SepFormer in the non-causal modality only.
SkiM also uses fewer MACs/s than the RE-SepFormer (3.7G versus 6.3G).
However, as we will show in the next section, this has no impact on latency.
Remarkably, even when handling long sequences, RE-SepFormer matches SkiM in terms of memory usage and inference speed.

\begin{table}[t]
\vspace{-0.4cm}
\centering
\caption{Comparison of RE-SepFormer and SkiM on WSJ0-2Mix and WHAM!.} 
\label{table:resepvsskim}
\resizebox{8.7cm}{!}{
\begin{tabular}{c c |c c|c c c c}
\textbf{} & \textbf{} & \multicolumn{2}{c|}{\textbf{Causal}} & \multicolumn{2}{c}{\textbf{Non-Causal}}  \\ 

\textbf{Model} & \textbf{Dataset} & \textbf{SI-SNRi (dB)}  & \textbf{SDRi (dB)} & \textbf{SI-SNRi (dB)} & \textbf{SDRi (dB)} \\
\hline  
SkiM & WSJ0-2Mix & 13.2  & 13.5  & 18.1 & 18.3\\ \hline
RE-SepFormer & WSJ0-2Mix & \textbf{14.2}  & \textbf{14.5} & \textbf{18.6}  & \textbf{18.9} \\ \hline 
SkiM & WHAM! & 10.6  & 11.0  & \textbf{14.5} & \textbf{14.8}\\ \hline
RE-SepFormer & WHAM! & \textbf{11.3}  & \textbf{11.7} & 14.1  & 14.4 \\ \hline
\end{tabular}
}
\vspace{-0.4cm}
\end{table}

\subsection{Speed and Memory Utilization}
\label{subsec:resourcecomparison}
\begin{figure}[t]
    \centering
    \begin{subfigure}{.235\textwidth}
      \includegraphics[width=1.0\linewidth]{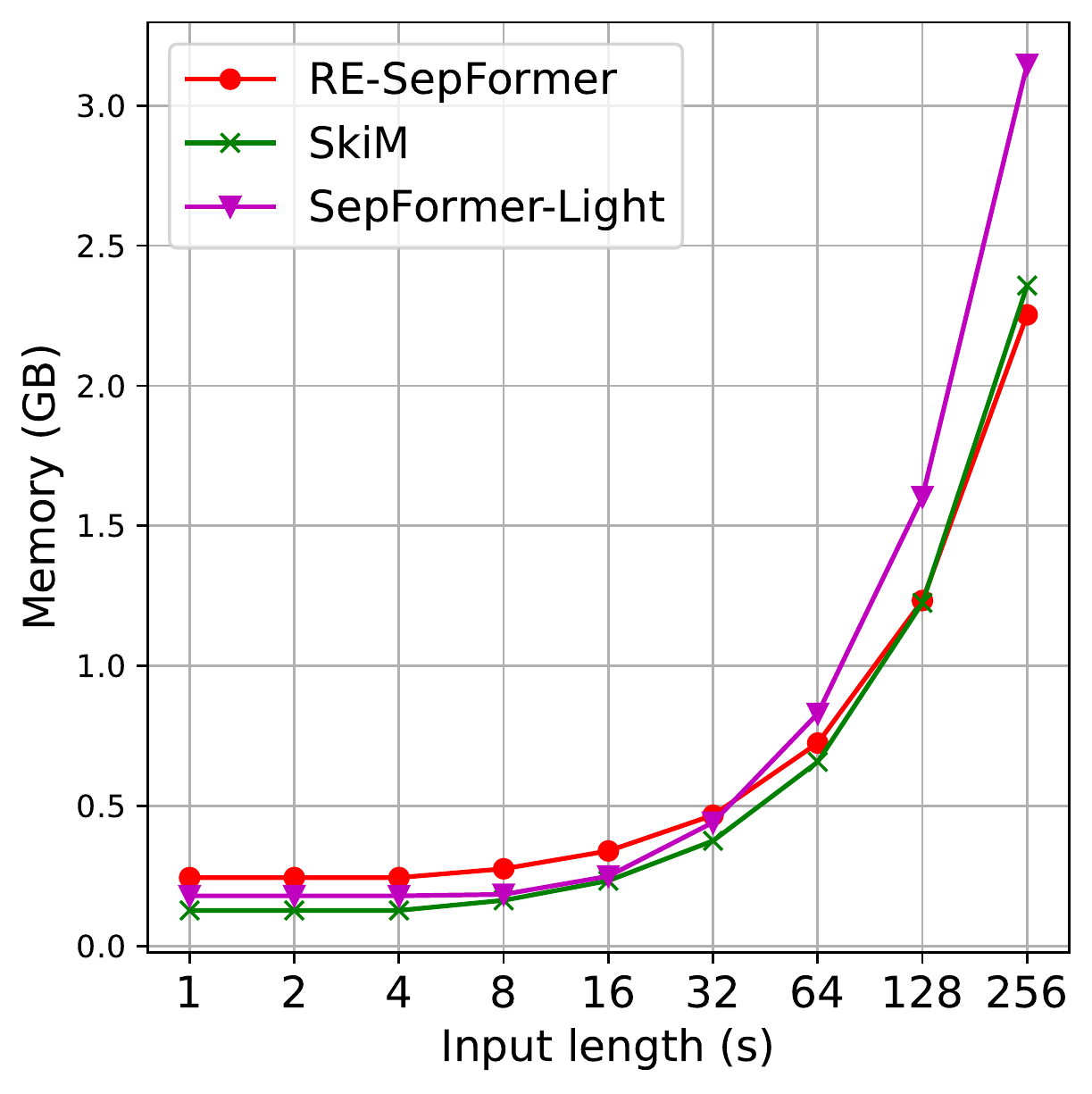}
    \end{subfigure}
    \begin{subfigure}{.235\textwidth}
      \includegraphics[width=1.0\linewidth]{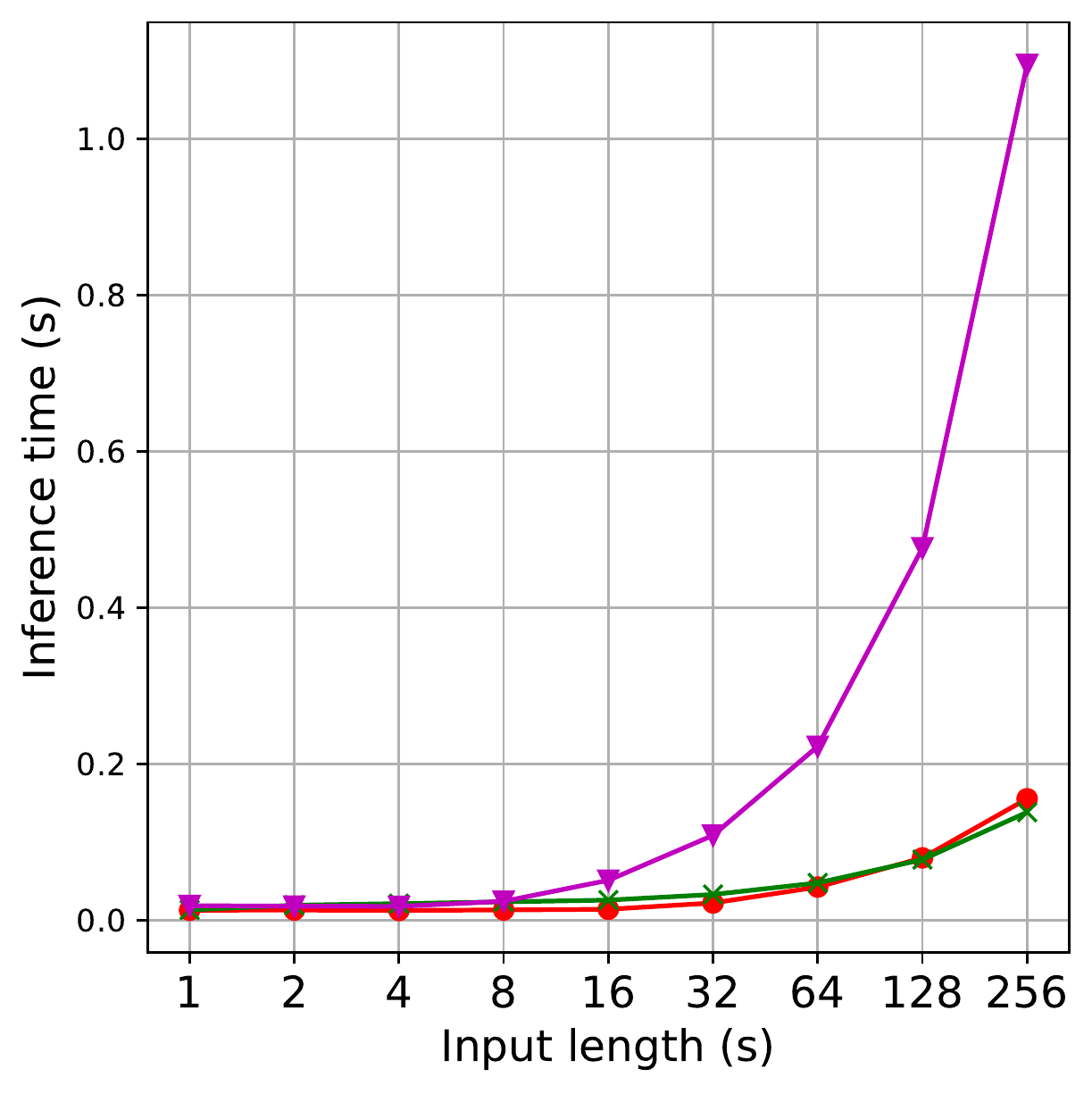}
    \end{subfigure}
    \vspace{-.3cm}
\caption{
Memory in GB (left panel) and inference time in seconds (right panel) comparison of RE-SepFormer, SkiM and SepFormer-Light. The x-axis in both panels shows the length of the input signal in seconds (8 kHz sampling rate).\vspace{-0.3cm}}
\label{fig:resource-comparison}
\end{figure}

In \cref{fig:resource-comparison}, we compare the memory usage (left) and the inference time (right) of the RE-SepFormer, SkiM, and SepFormer. 
For a more meaningful comparison, we use a SepFormer (denoted as SepFormer-Light) with a reduced number of parameters (i.e., 6.4 M). This experiment has been conducted on an NVIDIA A100 GPU considering different input lengths (ranging from 1 to 256 seconds).

The RE-SepFormer turned out to scale significantly better than the SepFormer-Light due to the summary representation. The most impressive result is the inference time observed when feeding the RE-SepFormer with long sequences. For an input of 256 seconds, the RE-SepFormer is 7x faster than SepFormer-Light.
Furthermore, it results in a memory usage reduction of up to 28\% for long sequences.\looseness=-1

It is worth noting that the RE-SepFormer is composed of self-attention blocks that consist of feed-forward layers. This feature makes the overall architecture highly parallelizable. In contrast, SkiM is mainly composed of RNN (LSTM) layers, which require sequential processing.
This potentially explains why it is not faster than RE-SepFormer, despite using only 60\% of the MACs.

\subsection{Comparison with Other Models}
In \cref{table:WSJ2Mix-noncausal}, we compare the performance of RE-SepFormer with a wide range of models from the literature (WSJ0-2Mix, non-causal setting). Despite its efficiency, we observe that RE-SepFormer achieves competitive performance. For example, RE-SepFormer performs comparably to Dual-Path RNN, while being significantly more efficient in terms of MACs.
The RE-SepFormer outperforms the popular Conv-TasNet, and SuDoRM-RF models. We also compare against popular efficient Transformer architectures (i.e., Reformer and Longformer) applied without chunking \cite{subakan2022onusing}. The RE-SepFormer outperforms aforementioned efficient Transformers as well.

\subsection{Ablation Studies}
To assess the relative importance of each component to the overall performance, we conduct the following ablation studies: (1) we decrease to 4 the number of layers in the IntraTransformer and Memory Transformer modules; (2) we reduce to 512 the dimension of the positional feed-forward layers in the IntraTransformer and Memory Transformer modules; (3) we combine all the previous ablations.

\cref{table:ablation} shows the results. Modifications to the IntraTransformers lead to the most significant performance drop together with the most substantial reduction in the number of parameters and MACs. In particular, halving the number of layers has the largest impact.
On the contrary, when we perform ablations on the Memory Transformer, we observe minimal effects on both the performance and MACs, despite a moderate reduction in the number of parameters.

It is worth noting that when we combine all the ablations, we still achieve an acceptable SDRi, surpassing methods like TasNet, SignPredictionNet and Conv-TasNet, while utilizing significantly fewer parameters and MACs. This further highlights the suitability of RE-SepFormer for small-footprint devices.

\section{Conclusions}
In this paper, we proposed the RE-SepFormer, which is a contribution towards more efficient speech separation with Transformers. 
The RE-SepFormer uses non-overlapping blocks and relies on compact latent summaries calculated from each chunk rather than attending all the time steps. 
Our experiments, conducted on the WSJ0-2Mix and WHAM! datasets in both causal and non-causal settings, show that the RE-SepFormer achieves an SDRi of 18.9 dB on WSJ0-2Mix and 14.4 dB on WHAM!.
Compared to the SepFormer, it employs more than 3x fewer parameters with a 11x reduction of the MACs. The model is mainly composed of feed-forward layers, and it is thus highly parallelizable. As a result, it scales significantly better than the previous Transformer-based architectures in terms of memory and inference time, making it more suitable for processing long mixtures.
This feature makes the RE-SepFormer particularly suitable for real-time low-latency speech separation on small-footprint devices such as GPU-equipped smartphones or laptops.\looseness=-1

\begin{table}[t]
\centering
\small
\caption{Best results on the WSJ0-2Mix dataset (test-set) for non-causal models.} 
\label{table:WSJ2Mix-noncausal}
\resizebox{8.7cm}{!}{
\begin{tabular}{l|c|c|c|c}
\textbf{Model} & \textbf{SI-SNRi (dB)} & \textbf{SDRi (dB)} & \textbf{\#Params (M)} & \textbf{GMACs/s} \\
\hline 
TasNet \cite{luo2017tasnet} & 10.8 & 11.1 & - & -  \\ \hline
SignPredictionNet \cite{wang2018deep} & 15.3 & 15.6 & 55.2 & - \\ \hline
Conv-TasNet \cite{luo2018convtasnet} & 15.3  & 15.6 & 5.1 & 3.2 \cite{skim} \\ \hline
Two-Step CTN \cite{tzinis2019twostep} & 16.1 & - & 8.6 & - \\ \hline
MGST \cite{zhao2020MSGtransformer} & 17.0 & 17.3 & - &  - \\ \hline
DeepCASA \cite{liu2019divide} & 17.7 & 18.0 & 12.8 & - \\ \hline
FurcaNeXt \cite{shi2019furcanext} & - & 18.4 & 51.4 & - \\ \hline
Dual-Path RNN \cite{luo2020dualpath} & 18.8 & 19.0 & 2.6 & 38.9 \cite{skim} \\ \hline
SuDoRM-RF \cite{tzinis2020sudo} &  17.0 & - & 2.6  & -  \\ \hline
DPTNet \cite{dptn} & 20.2 & 20.6 & 2.6 & - \\  \hline
SkiM \cite{skim} + DM & 18.2 & 18.4 & 14.5 & 3.7\\  \hline
SepFormer \cite{subakan2020attention} + DM& 22.3 & 22.4 & 25.7 & 69.6 \\ \hline
SepFormer Light + DM & 20.0 & 20.2 & 6.4 & 17.5 \\ \hline
Reformer \cite{subakan2022onusing} + DM & 16.7 & 16.9 & 12.0 & 12.4 \\ \hline
Longformer \cite{subakan2022onusing} + DM &  13.1 & 13.4 & 15.1 & 12.3 \\ \hline
SepIt + DM \cite{sepit} & 22.4 & - & 4.6 & - \\ \hline
TFPSNet \cite{yang2022tfpsnet} & 21.1 & 21.3 & 2.7 & 29.6 \cite{wang2023tfgridnet} \\ \hline
TF-GridNet \cite{wang2023tfgridnet} & - & 23.6  & 14.5 & 231.1 \cite{wang2023tfgridnet} \\ \hline
MossFormer + DM \cite{Zhao2023MossFormerPT}  &  22.8 & - & 42.1 & 42.7 \\
\hline \hline
{RE-SepFormer } + DM & 18.6  & 18.9 & 8.0 & 6.3 \\ 
\hline 
\end{tabular}
}
\end{table}

\begin{table}[t]
\centering
\caption{Ablation studies on the number of intra/memory layers and the intra/memory positional feed-forward layer dimension (d$_{\text{ff}}$), non-causal setting.}
\label{table:ablation}
\resizebox{8.7cm}{!}{
\begin{tabular}{c|c|c|c|c|c|c}
\textbf{\#Intra} & \textbf{Intra d$_{\text{ff}}$}  & \textbf{\#Memory} & \textbf{Memory d$_{\text{ff}}$} & \textbf{SDRi (dB)} & \textbf{\#Params (M)} & \textbf{GMACs/s} \\
\hline  
8 & 1024  & 8  & 1024  & 18.9 & 8.0 & 6.3  \\ \hline
4 & 1024  & 8  & 1024  &  16.4 & 5.3 & 3.2   \\ \hline
8 & 512   & 8  & 1024  & 18.3 & 5.8 & 4.0  \\ \hline
8 & 1024  & 4  & 1024  &  18.5 & 6.6 & 6.2  \\ \hline
8 & 1024  & 8  & 512  &  18.7 &  6.9 & 6.2  \\ \hline
4 & 512  & 4  & 512  & 16.5 & 2.4 & 2.0 \\ \hline
\end{tabular}
    }
\end{table}

\bibliographystyle{IEEEbib-abbr}
\bibliography{refs}

\end{document}